\documentclass[fleqn,twoside]{article}
\usepackage{gc}

\heads{Tomohiro Inagaki}
      {Dynamical Symmetry Breaking in Randall-Sundrum background}

\begin{document}
\twocolumn[
\Arthead{9}{2003}{1-2 (33-34)}{41}{44}

\Title{DYNAMICAL SYMMETRY BREAKING  \yy
        IN RANDALL-SUNDRUM BACKGROUND}

   \Author{Tomohiro Inagaki\foom 1} 
          {Information Media Center, Hiroshima University, Kagamiyama 1-7-1, 
Higashi-Hiroshima, 739-8521, Japan} 

\Abstract
{
Characteristic features of dynamical symmetry breaking are investigated in 
Randall-Sundrum (RS) background.
To study the influence of bulk fermion or bulk gauge field I consider a 
five-dimensional four-fermion interaction model and a non-Abelian gauge field 
theory as simple models where the chiral symmetry is broken down dynamically. 
Using a four-dimensional Lagrangian induced from the five-dimensional theory, 
I calculate the dynamically generated fermion mass and show how the extra 
dimension affects the chiral phase transition in the  RS background.
}

]  %

\email 1 {inagaki@hiroshima-u.ac.jp}

\section{Introduction}
The idea of compact extra dimensions has been proposed to solve the 
cosmological constant problem, why our Universe seems flat even if we have 
a cosmological constant appearing from a vacuum expectation value 
\cite{Rubakov:83bz}, and the hierarchy problem between the Planck scale 
and the electroweak scale \cite{Antoniadis:90ew}, 
\cite{Arkani-Hamed:98rs}. 
A brane is a four-dimensional object embedded in a higher dimensional 
space-time, called bulk. In the original idea of the brane world scenario the 
standard model particles are localized on the Minkowski brane, while the 
graviton can propagate in extra dimensions, the bulk.

A dynamical mechanism of electroweak symmetry breaking has been re-examined 
in compact extra dimensions in Refs. \cite{Dobrescu:98dg}, 
\cite{Cheng:99bg}, \cite{Abe:00ny}, 
\cite{Arkani-Hamed:00hv}, \cite{Gusynin:02cu}. The finite size effect 
and the boundary condition can dramatically change the phase structure of 
dynamical symmetry breaking \cite{Ishikawa:uu}. It has been pointed out that 
exchange of Kaluza-Klein (KK) modes enhances dynamical symmetry breaking. 
A possibility of the top quark condensation is revived along with the brane 
world scenario \cite{Hashimoto:00uk}. 

To avoid a fine tuning of the radius of compact extra dimensions, Randall 
and Sundrum introduced a 5D space-time with a curved extra dimension 
\cite{Randall:99ee}. In their model it is assumed that the fifth dimension 
$\theta$ is compactified on an orbifold, $S^1/Z_2$, and two Minkowski branes 
exist at the orbifold fixed points, $\theta=0$ and $\pi$. It was found that 
this space-time satisfies the Einstein equation for special bulk and brane 
cosmological constants. The RS background is a 5D anti-de-Sitter space 
described by the metric,
\begin{equation}
g^{\mu\nu}=e^{-2kr|\theta|}\eta_{\mu\nu}dx^{\mu}dx^{\nu}+r^2d\theta^2 .
\end{equation}
The exponential factor $e^{-2kr|\theta|}$ is called a ``warp factor''.

The RS brane world has two Minkowski branes in five-dimensional 
negative-curvature space-time, $AdS^5$. It has been known that the 
space-time curvature has an important effect on the phase structure of 
dynamical symmetry breaking \cite{Inagaki:93ya}, \cite{Inagaki:95jp},
\cite{Ishikawa:yx}, \cite{Rius:01dd}. 
Especially in negative-curvature space-time it has been shown that a chiral 
symmetry is always broken down for models with four-fermion interactions 
if the space-time dimension is less than four \cite{Inagaki:95bk}, 
\cite{Inagaki:97kz}. 

In the RS background, the effective Planck scale $M_{pl}$ , i.e. mass 
scale for gravity, is given by the ration 
\begin{equation}
{M_{pl}}^2\sim\frac{M^3}{k}(1-e^{-2kr\pi})\sim\frac{M^3}{k} ,
\end{equation}
where $M$ is the fundamental scale in the bulk and $k$ is the curvature.
On the other hand, the mass scale $M_{phys}$ on the $\theta=\pi$ brane 
is suppressed by the warp factor,
\begin{equation}
M_{phys}=M e^{-kr\pi} .
\end{equation}
For $k\sim 11$, it is able to consider the mass scale
\begin{equation}
\left\{
\begin{array}{l}
M\sim k\sim \mbox{O}(M_{pl}) , \\
M_{phys}\sim \mbox{O}(M_{EW}) .
\end{array}
\right.
\end{equation}
Therefore the electroweak mass scale can be naturally realized from only the 
Planck scale without introducing some large number. This is the most important 
mechanism of the RS model. I want to realize this mechanism dynamically.
For this purpose we launched a plan to evaluated some models of dynamical 
symmetry breaking in the RS brane world \cite{Abe:01yb}, \cite{Abe:01yi},
\cite{Abe:02yb}.

In the present paper we consider a bulk standard model where the standard 
model particles also propagate in the extra dimensions. In the bulk standard 
model the KK excitation modes of the standard model particles appear on the 
brane. These KK modes may affect some of low energy phenomena on the brane. 
One of the most interesting phenomena is found in a spontaneous electroweak 
symmetry breaking. 
It is expected that a negative curvature enhances the symmetry breaking. 
I discuss the contribution from the KK modes to the dynamical symmetry 
breaking.

First I will introduce the bulk fermion in the RS background. I show that a 
low fermion mass is generated dynamically. Next I analyze a bulk gauge 
theory by using the improved ladder SD equation. Finally I give the 
concluding remarks. 

\section{Bulk four-fermion interaction model}
As a simplest example of dynamical symmetry breaking, we consider a
four-fermion interaction model with bulk fermions,
\begin{eqnarray}
{\cal L}^{5D}&=&\sqrt{-G}\left[
\bar{\psi}_1^{5D}i\partial\!\!\!/ \psi_1^{5D}
+\bar{\psi}_2^{5D}i\partial\!\!\!/ \psi_2^{5D}
\right. \nonumber \\
&& \left.  \mbox{\hspace*{6ex}}
+\lambda(\bar{\psi}_1^{5D}\psi_2^{5D})(\bar{\psi}_2^{5D}\psi_1^{5D}) 
\right] ,
\label{L:ff}
\end{eqnarray}
where we assume the existence of two kinds of bulk fermions with different 
parity,
\begin{equation}
\left\{
\begin{array}{l}
\psi_1^{5D}(x,\theta)= \gamma_5 \psi_1^{5D}(x,-\theta) ,\\
\psi_2^{5D}(x,\theta)= -\gamma_5 \psi_2^{5D}(x,-\theta) .
\end{array}
\right.
\end{equation}
Two kinds of fermions are necessary for constructing Dirac mass terms for 
bulk fermions in the induced 4D theory \cite{Chang:99nh}.

The bulk fermion is expanded into KK modes,
\begin{equation}
\psi^{5D}(x,\theta)=\sum_{n=0}^{\infty}\psi_{R}^{(n)}(x)g_{R}^{(n)}(\theta)
+\psi_{L}^{(n)}(x)g_{L}^{(n)}(\theta) ,
\label{KK:psi}
\end{equation}
where $g_{L}^{(n)}$ and $g_{R}^{(n)}$ are left and right mode functions 
which satisfy
\begin{equation}
\left\{
\begin{array}{l}
\displaystyle \int d\theta e^{-3kr|\theta|} g_{L}^{(n)}(\theta) 
g_{L}^{(n)}(\theta) = \delta_{mn} , \\[1.4mm]
\displaystyle \int d\theta e^{-3kr|\theta|} g_{R}^{(n)}(\theta) 
g_{R}^{(n)}(\theta) = \delta_{mn} .
\end{array}
\right.
\end{equation}
Substituting Eq.(\ref{KK:psi}) into (\ref{L:ff}) and applying the auxiliary 
field method, the Lagrangian reduces to
\begin{eqnarray}
&&{\cal L}=\bar{\psi}^{(0)}_{1R}i\gamma^{\mu}\partial_{\mu}\psi^{(0)}_{1R}
+\bar{\psi}^{(0)}_{2L}i\gamma^{\mu}\partial_{\mu}\psi^{(0)}_{2L}
\nonumber \\
&&+\sum_{1\leq n}\left[
\bar{\psi}^{(n)}_{1R}i\gamma^{\mu}\partial_{\mu}\psi^{(n)}_{1R}
+\bar{\psi}^{(n)}_{1L}i\gamma^{\mu}\partial_{\mu}\psi^{(n)}_{1L}
\right.
\nonumber \\
&&\left.
\bar{\psi}^{(n)}_{2R}i\gamma^{\mu}\partial_{\mu}\psi^{(n)}_{2R}
+\bar{\psi}^{(n)}_{2L}i\gamma^{\mu}\partial_{\mu}\psi^{(n)}_{2L}
\right]
\nonumber \\
&&+\sum_{m,n=0}^{\infty}
\left(\bar{\psi}^{(m)}_{1R}\ \bar{\psi}^{(m)}_{2R}\ 
\bar{\psi}^{(m)}_{1L}\ \bar{\psi}^{(m)}_{2L}\right)
M
\left(
\begin{array}{c}
\psi^{(n)}_{1R} \\
\psi^{(n)}_{2R} \\
\psi^{(n)}_{1L} \\
\psi^{(n)}_{2L}
\end{array}
\right)
\nonumber \\
&&-\int d\theta r \sqrt{-G} \frac{|\sigma|^2}{\lambda} .
\end{eqnarray}
Integrating over the fifth direction $\theta$, we obtain the induced 
4D Lagrangian.
Starting from the induced 4D Lagrangian, we calculate the effective potential. 
The vacuum expectation value of the composite operator $\bar{\psi}\psi$ is 
determined by observing the minimum of the effective potential. 
It is proportional to the dynamical mass of the fermion. 
Here we suppose a specific form for the vacuum expectation value and
restrict ourselves to some specific cases to integrate over the fifth 
direction $\theta$ analytically.

There is no translational invariance along the $\theta$ direction in the RS 
background.
The vacuum expectation vale $\langle\bar{\psi}\psi\rangle$ is not necessary 
constant. We suppose the following form for the vacuum expectation value 
of the composite operator,
\begin{equation}
\langle\bar{\psi}\psi\rangle = \frac{1}{\lambda}v e^{kr\theta} ,
\end{equation}
and evaluate the effective potential. In this case the minimal vale of the 
effective potential is smaller than the constant 
$\langle\bar{\psi}\psi\rangle$ if the number of KK mode is not so small. 
I notice that the warp suppression in the fermion mass term is canceled out 
by the factor $e^{kr\theta}$.

There is a second order phase transition as $\lambda$ increased. The chiral 
symmetry is broken above the critical coupling $\lambda_c$. 
It is found that the critical coupling $\lambda_c$ is warp-suppressed. 
\begin{equation}
\lambda_c \sim  \frac{2\pi^3}{\Lambda^3} e^{-kr\theta} ,
\end{equation}
where $\Lambda$ is the cut-off scale.
The vacuum expectation vale $\langle\bar{\psi}\psi\rangle$ is generated 
dynamically even for an extremely small coupling. But it is not suppressed 
by the warp factor. The fermion field acquires a Planck scale mass 
correction.

On the other hand, the dynamical fermion mass is given by
\begin{equation}
m_n^{\pm}\sim \left| v \pm \frac{nk\pi}{e^{k\pi r}-1} \right| , \ \ 
n=0, 1, 2, \cdots ,
\end{equation}
for each KK mode.
The lightest fermion mass is smaller than the mass scale $M_{phys}$ on the 
$\theta=\pi$ brane even if the vacuum expectation value $v$ develops a value 
near the Planck scale \cite{prepare:02}. 
Therefore the low mass fermion is generated dynamically in the bulk 
four-fermion interaction model.

\section{Bulk gauge theory}
Next we assume the existence of five-dimensional bulk gauge fields, 
while the fermion fields are confined to the brane. 
The Lagrangian of a bulk gauge theory is given by
\begin{eqnarray}
{\cal L}^{5D}&=&\sqrt{-G}\left[-\frac{1}{4}{F^{5D}}^{MN}{F^{5D}}_{MN}
\right. \nonumber \\
&&\left.
\mbox{\hspace*{6ex}}
+\bar{\psi}(i\partial\!\!\!/ +g{A\!\!\!/}^{5D})\psi\delta(\theta-\theta^*)
\right] ,
\end{eqnarray}
where $\theta^*$ describes the brane fixed points.
The KK mode expansion for the gauge field is given by
\begin{eqnarray}
{A^{5D}}_{\mu}(x,\theta)&=&
\sum_{n=0}^{\infty} {A^{(n)}}_{\mu}(x)\chi_{n}(\theta) 
\nonumber \\
&&+ \sum_{n=1}^{\infty} {\tilde{A}^{(n)}}_{\mu}(x)\tilde{\chi}_{n}(\theta) , 
\end{eqnarray}
where $\chi_n$ and $\tilde{\chi}_n$ are the $Z_2$-even and odd
mode functions which satisfy
\begin{equation}
\left\{
\begin{array}{l}
\displaystyle
\frac{1}{2\pi} \int d \theta \chi_{m}(\theta) \chi_{n}(\theta) = \delta_{mn} ,
\\
\displaystyle
\frac{1}{2\pi} \int d \theta \tilde{\chi}_{m}(\theta) \tilde{\chi}_{n}(\theta) 
= \delta_{mn} .
\end{array}
\right.
\end{equation}
Applying the KK mode expansion and integrating over the extra direction, 
we obtain the induced 4D Lagrangian on the brane \cite{Abe:01yi},
\cite{Abe:02yb},
\begin{eqnarray}
&&{\cal L}_{eff}=\sum_{n=0}^{\infty}
\left[-\frac{1}{4}{F^{(n)}}^{\mu\nu}F^{(n)}_{\mu\nu}
+\frac{1}{2}{M_n}^2 {A^{(n)}}^\mu A^{(n)}_{\mu}\right]\nonumber \\
&&+\sum_{n=1}^{\infty}\left[
-\frac{1}{4}{\tilde{F}^{(n)}}{}^{\mu\nu}\tilde{F}^{(n)}_{\mu\nu}
+\frac{1}{2}{M_n}^2 {\tilde{A}^{(n)}}{}^\mu \tilde{A}^{(n)}_{\mu}\right]
\nonumber \\
&&+\sum_{n=0}^{\infty}g\chi_{n}(\theta)\bar{\psi}({A\!\!\!/}^{(n)})\psi
\nonumber \\
&&+\sum_{n=1}^{\infty}g\tilde{\chi}_{n}(\theta)\bar{\psi}
({\tilde{A}\!\!\!/}^{(n)})\psi ,
\label{4D:L:g}
\end{eqnarray}
where we take the $A_{4}=0$ gauge.
Since the KK-excited gauge field is localized at $\theta=\pi$, the 4D 
effective coupling between KK-excited modes and the brane fermion is 
enhanced by $\chi_n$ \cite{Chang:99nh}. Thus an effective strong coupling 
is obtained for KK-excited modes. This strong coupling enhances the 
dynamical symmetry breaking.

Starting from the Lagrangian (\ref{4D:L:g}), we derive the Schwinger-Dyson 
(SD) equation for the full brane fermion propagator on the $\theta=\pi$ brane.
Here we consider the bulk SU(3) gauge theory, i.e., bulk QCD.
We replace the full gauge boson propagator and the full vertex function in 
the SD equation with the free propagator and the 1-loop running coupling 
constant. It is well-known improved ladder approximation \cite{Miransky:vj}, 
\cite{Higashijima:83gx}. 
The running coupling has a power law behavior in the truncated KK approach
\cite{Dienes:98vh}, \cite{Dienes:98vg}.
In Refs.~\cite{Randall:01gb} it is pointed out that the running is only
logarithmic in RS background. If the gauge coupling has no power law 
correction, the effect of the KK modes will be slightly enhanced.
Roughly speaking, the KK modes propagation enhances the dynamical symmetry 
breaking, while the power-law running suppresses it.

We numerically solved the improved ladder SD equation and found the behaviors 
of the fermion wave function and the mass function on the $\theta=\pi$ brane.
A second order phase transition takes place at the critical value of the 
coupling constant. It should be noted that the gauge-fixing parameter is 
chosen to satisfy the QED-like WT identity, i.e., the wave function correction 
disappears. In our numerical analysis the wave function correction is smaller 
than 0.7\%.
We found that the scale of the fermion mass function is near the scale on the 
brane. On $\theta=\pi$ brane the dynamical fermion mass is warp-suppressed.

We can calculate the decay constant for a composite scalar field by using 
the Pagels-Stokar formula \cite{Pagels:hd}, \cite{Aoki:90eq}. 
The scale of the decay constant is also near the mass scale on the brane.
We can calculate the same quantity in the 5D flat extra dimension. 
In this case the scale of the decay constant is located near the QCD scale
$\Lambda_{QCD}$. The effective strong coupling $g\chi_n$ enhances the 
dynamical symmetry breaking in the RS background. Therefore it is 
possible to obtain an electroweak-scale decay constant from 
bulk QCD \cite{Abe:02yb}.

\section{Summary}
Two models of dynamical symmetry breaking have been investigated in the RS 
background. We have studied a model where fermion  propagates in the extra 
dimension, i.e., a bulk fermion model, and a theory where a gauge boson 
propagates in the extra dimension,  i.e., a bulk gauge theory. 
Using the induced 4D theory, we have evaluated the dynamically generated 
fermion mass.

First we have considered the model with bulk fermions. Assuming the 
existence of two kinds of bulk fermions, we have calculated the 4D 
effective potential. The chiral symmetry is broken down through 
a second-order phase transition for $\lambda > \lambda_{cr}$.
It is natural to take the scale of $\langle{\bar{\psi}}^{5D}\psi^{5D}\rangle$ 
as the fundamental scale of the bulk. It is found that a low-mass fermion 
appears even if the vacuum expectation value, 
$\langle{\bar{\psi}}^{5D}\psi^{5D}\rangle$, is near the Planck scale.

Next we have studied bulk QCD. The bulk gauge boson does not decouple on the 
$\theta=\pi$ brane due to the effective strong coupling. It follows from the 
localization of KK modes. I analyze the effects of the KK mode propagation 
and power-law running. We have numerically solved the SD equation for the 
full brane fermion propagator. It is found that the dynamical mass of the 
brane fermion and the decay constant of the composite Nambu-Goldstone 
scalar are warp-suppressed from the fundamental scale at $\theta=\pi$.
The electroweak mass scale can be realized from only the Planck scale in the 
RS brane world due to the fermion and anti-fermion condensation. 
It is a dynamical realization of the Randall-Sundrum mechanism.

The models discussed here are only simple toy models. Some extension is 
necessary to construct a realistic model of electroweak symmetry breaking. 
Here we have not considered the Yukawa coupling for light fermions. We 
must introduce the origin of light fermion masses and explain the hierarchy 
in the Yukawa coupling.

\Acknow
{    
The main part of this paper is based on the works \cite{Abe:01yi},
\cite{Abe:02yb} and \cite{prepare:02} in collaboration 
with H.~Abe, K.~Fukazawa, Y.~Katsuki, T.~Muta and K.~Ohkura.
}


\small
\begin{thebibliography}{99}

\bibitem{Rubakov:83bz}
V.~A.~Rubakov and M.~E.~Shaposhnikov,
{\it Phys. Lett.\/} {\bf B 125}, 139 (1983). 

\bibitem{Antoniadis:90ew}
I.~Antoniadis,
{\it Phys. Lett.\/} {\bf B 246}, 377 (1990). 

\bibitem{Arkani-Hamed:98rs}
N.~Arkani-Hamed, S.~Dimopoulos and G.~R.~Dvali,
{\it Phys. Lett.\/} {\bf B 429}, 263 (1998). 

\bibitem{Dobrescu:98dg}
B.~A.~Dobrescu,
{\it Phys. Lett.\/} {\bf B 461}, 99 (1999). 

\bibitem{Cheng:99bg}
H.~C.~Cheng, B.~A.~Dobrescu and C.~T.~Hill,
{\it Nucl. Phys.\/} {\bf B 589}, 249 (1998). 

\bibitem{Abe:00ny}
H.~Abe, H.~Miguchi and T.~Muta,
{\it Mod. Phys. Lett.\/} {\bf A 15}, 445 (2000). 

\bibitem{Arkani-Hamed:00hv}
N.~Arkani-Hamed, H.~C.~Cheng, B.~A.~Dobrescu and L.~J.~Hall,
{\it Phys. Rev.\/} {\bf D 62}, 096006 (2000). 

\bibitem{Gusynin:02cu}
V.~Gusynin, M.~Hashimoto, M.~Tanabashi and K.~Yamawaki,
{\it Phys. Rev.\/} {\bf D 65}, 116008 (2002). 

\bibitem{Ishikawa:uu}
K.~Ishikawa, T.~Inagaki, K.~Yamamoto and K.~Fukazawa,
{\it Prog. Theor. Phys.\/} {\bf 99}, 237 (1998). 

\bibitem{Hashimoto:00uk}
M.~Hashimoto, M.~Tanabashi and K.~Yamawaki,
{\it Phys. Rev.\/} {\bf D 64}, 056003 (2001). 

\bibitem{Randall:99ee}
L.~Randall and R.~Sundrum,
{\it Phys. Rev. Lett\/} {\bf 83}, 3370 (1999). 

\bibitem{Inagaki:93ya}
T.~Inagaki, T.~Muta and S.~D.~Odintsov,
{\it Mod. Phys. Lett.\/} {\bf A 8}, 2117 (1993). 

\bibitem{Inagaki:95jp}
T.~Inagaki, S.~Mukaigawa and T.~Muta,
{\it Phys. Rev.\/} {\bf D 52}, 4267 (1995). 

\bibitem{Ishikawa:yx}
K.~Ishikawa, T.~Inagaki and T.~Muta,
{\it Mod. Phys. Lett.\/} {\bf A 11}, 939 (1996). 

\bibitem{Rius:01dd}
N.~Rius and V.~Sanz,
{\it Phys. Rev.\/} {\bf D 64}, 075006 (2001). 

\bibitem{Inagaki:95bk}
T.~Inagaki,
{\it Int. J. Mod. Phys.\/} {\bf A 11}, 4561 (1996). 

\bibitem{Inagaki:97kz}
T.~Inagaki, T.~Muta and S.~D.~Odintsov,
{\it Prog. Theor. Phys. Suppl.\/} {\bf 127}, 93 (1997). 

\bibitem{Abe:01yb}
H.~Abe, T.~Inagaki and T.~Muta, {\it in:\/} ``Fluctuating Paths and Fields '', ed. W. Janke, A. Pelster, H.-J. Schmidt, M. Bachmann, World Scientific, 2001.

\bibitem{Abe:01yi}
H.~Abe, K.~Fukazawa and T.~Inagaki,
{\it Prog. Theor. Phys.\/} {\bf 107}, 1047 (2002). 

\bibitem{Abe:02yb}
H.~Abe and T.~Inagaki,
{\it Phys. Rev.\/} {\bf D 66}, 085001 (2002). 

\bibitem{Chang:99nh}
S.~Chang, J.~Hisano, H.~Nakano, N.~Okada and M.~Yamaguchi,
{\it Phys. Rev.\/} {\bf D 62}, 084025 (2000). 

\bibitem{prepare:02} 
K.~Fukazawa, T.~Inagaki, Y.~Katsuki, T.~Muta and K.~Ohkura,
hep-ph/0308022.

\bibitem{Miransky:vj}
V.~A.~Miransky,
{\it Sov. J. Nucl. Phys.\/} {\bf 38}, 280 (1983), 
{\it Yad. Fiz.\/} {\bf 38}, 468 (1983). 

\bibitem{Higashijima:83gx}
K.~Higashijima,
{\it Phys. Rev.\/} {\bf D 29}, 1228 (1984). 

\bibitem{Dienes:98vh}
K.~R.~Dienes, E.~Dudas and T.~Gherghetta,
{\it Phys. Lett.\/} {\bf B 436}, 55 (1998). 

\bibitem{Dienes:98vg}
K.~R.~Dienes, E.~Dudas and T.~Gherghetta,
{\it Nucl. Phys.\/} {\bf B 537}, 47 (1999). 

\bibitem{Randall:01gb}
L.~Randall and M.~D.~Schwartz,
{\it JHEP\/} {\bf 0111}, 003 (2001). 

\bibitem{Pagels:hd}
H.~Pagels and S.~Stokar,
{\it Phys. Rev.\/} {\bf D 20}, 2947 (1979). 

\bibitem{Aoki:90eq}
K.~I.~Aoki, M.~Bando, T.~Kugo, M.~G.~Mitchard and H.~Nakatani,
{\it Prog. Theor. Phys.\/} {\bf 84}, 683 (1990). 

\end{thebibliography}
\end{document}